\documentclass[10pt,letterpaper]{article}
\usepackage[utf8x]{inputenc}
\usepackage[top=0.85in,left=2.75in,footskip=0.75in]{geometry}
\usepackage{amsmath,amssymb,listings}
\usepackage{changepage}
\usepackage{cite}
\usepackage[colorlinks=true]{hyperref}
\usepackage{microtype}
\DisableLigatures[f]{encoding = *, family = * }
\usepackage[dvipsnames]{xcolor}

\newlength\savedwidth

\usepackage[aboveskip=1pt,labelfont=bf,labelsep=period,justification=raggedright,singlelinecheck=off]{caption}

\makeatletter
\renewcommand{\@biblabel}[1]{\quad#1.}
\makeatother

\usepackage{lastpage,fancyhdr,graphicx}
\usepackage{epstopdf}
\fancyheadoffset[L]{2.25in}
\fancyfootoffset[L]{2.25in}
\newcommand{\ie}{\emph{i.e.}}               
\newcommand{\abs}[1]{\lvert #1\rvert}       
\begin{document}
\vspace*{0.2in}

\begin{flushleft}
{\Large
\textbf\newline{ProtRank: Bypassing the imputation of missing values in differential expression analysis of proteomic data}
}
\newline
\\
Matúš Medo\textsuperscript{1,2,3,4*},
Daniel M. Aebersold\textsuperscript{1,2},
Michaela Medová\textsuperscript{1,2}\\

\bigskip
\textbf{1} Department of Radiation Oncology, Inselspital, Bern University Hospital and University of Bern, 3010 Bern, Switzerland\\
\textbf{2} Department for BioMedical Research, Inselspital, Bern University Hospital and University of Bern, Bern, Switzerland\\
\textbf{3} Institute of Fundamental and Frontier Sciences, University of Electronic Science and Technology of China, Chengdu 610054, PR China\\
\textbf{4} Department of Physics, University of Fribourg, 1700 Fribourg, Switzerland\\
\bigskip

* matus.medo@unifr.ch
\end{flushleft}

\section*{Abstract}

\subsubsection*{Background}
Data from discovery proteomic and phosphoproteomic experiments typically include missing values that correspond to proteins that have not been identified in the analyzed sample. Replacing the missing values with random numbers, a process known as ``imputation'', avoids apparent infinite fold-change values. However, the procedure comes at a cost: Imputing a large number of missing values has the potential to significantly impact the results of the subsequent differential expression analysis.

\subsubsection*{Results}
We propose a method that identifies differentially expressed proteins by ranking their observed changes with respect to the changes observed for other proteins. Missing values are taken into account by this method directly, without the need to impute them. We illustrate the performance of the new method on two distinct datasets and show that it is robust to missing values and, at the same time, provides results that are otherwise similar to those obtained with edgeR which is a state-of-art differential expression analysis method.

\subsubsection*{Conclusion}
The new method for the differential expression analysis of proteomic data is available as an easy to use Python package.

\subsubsection*{Keywords}
Proteomics; Differential expression analysis; Ranking; Imputation; Significance

\section*{Background}
The recent availability of high-resolution omic measurements has called for the creation of statistical methods and tools to analyze the resulting data~\cite{anders2010differential,robinson2010edger,law2014voom,anders2015htseq}. Proteomics, a large-scale analysis of proteins in biomaterials such as cells or plasma, in particular, can help elucidate molecular mechanisms of disease, aging, and effects of the environment~\cite{vizcaino2014proteomexchange}. Expression proteomics, quantitative study of protein expression between samples that differ by some variable, is used to identify novel proteins in signal transduction or disease-specific proteins~\cite{larance2015multidimensional,tyanova2016perseus}. The application of proteomic technologies to clinical specimens has the potential to revolutionize the treatment of many diseases: From biomarker discovery and validation to personalized therapies, proteomic techniques allow a greater understanding of the dynamic processes involved in disease, increasing the power of prediction, diagnosis, and prognosis~\cite{latterich2008proteomics,guest2013proteomics,ebhardt2015applications,frantzi2018clinical}. Detailed measurements of protein levels allow for characterizing protein modifications and identifying the targets of drugs~\cite{graves2002molecular}. The quantitative study of protein expression between samples that differ by some variable is known as expression proteomics. In this approach, protein expression of the entire proteome or its subproteomes between samples can be compared. This can help identify novel proteins in signal transduction or identify disease-specific proteins.

However, data from proteomic and phosphoproteomic experiments are not error-free. Of various measurement errors, missing values are particularly severe. They arise when signals from some proteins are not detected by the instrument. Due to the technical setup of measurements, proteomic data often contain a considerable fraction of missing (zero) values. To avoid mathematical difficulties (such as infinite or very large logarithmic fold changes in pairwise comparisons involving a missing value), missing values are typically removed by a process which is referred to as \emph{imputation}: All missing values are replaced by samples from a given distribution~\cite{tyanova2016perseus}. Besides the need to choose the distribution's parameters, the often-ignored drawback of value imputation is that it has the potential to distort the analysis results. We use two real datasets to show that imputation indeed significantly alters the analysis results. We propose a method for differential expression analysis of proteomic data where missing values can be taken into account naturally, without the need to replace them by random numbers.

To demonstrate that the new method performs well, we use transcriptomic data where missing values do not pose a problem. Standard methods for differential expression analysis, such as the edgeR package~\cite{robinson2010edger} which performs well in many cases~\cite{soneson2013comparison}, can be therefore used to produce reliable results on transcriptomic data. We use the thus-obtained results as a benchmark with which we compare the results produced by the newly proposed method. Subsequently, we introduce artificial missing values in the analyzed data to demonstrate that the new method is significantly more robust to the presence of missing values than edgeR. We conclude by analyzing phosphoproteomic data where problematic missing values occur naturally. An implementation of the new method in Python is available at \url{https://github.com/8medom/ProtRank} (see \nameref{sec:PR_usage} for a brief usage description).

\section*{Results}

\subsection*{Missing values in proteomic data and their impact}
\label{sec:missing_values}
The simplest way how to deal with missing values is to ignore them and analyze only the proteins that have no missing values at all. This is problematic for two reasons. First, proteomic data typically involve a large fraction of missing values and the proteins that have no missing values can be thus correspondingly scarce. In the case of the phosphoproteomic data analyzed later (see \nameref{sec:Jonas_data} for details), for example, the overall fraction of missing values is 43.6\% and only 37.9\% of all proteins have no missing values. The data on the remaining 62.1\% of proteins would be wasted in this case. Second, the missing values can contain important information: A protein can be absent in the results not due to a measurement error but because of actual biological processes---application of an inhibitive treatment, for example. For this reason, we need an approach that can analyze proteomic data where missing values are still present.

\begin{figure}
\centering
\includegraphics[scale=0.7]{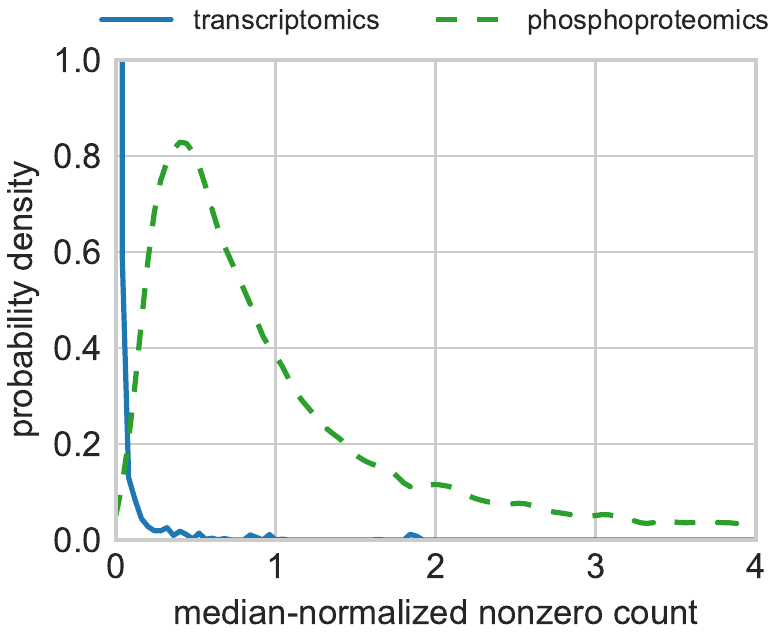}
\caption{\textbf{The distribution of positive counts in pairwise comparisons where the other count is zero.} We compare here transcriptomic and phosphoproteomic data. To make the two datasets directly comparable, the positive counts are scaled by the datasets' respective median counts.}
\label{fig:data_properies}
\end{figure}

A comparison between expression values in different samples is usually based on logarithmic fold change values. Denoting the counts of gene $g$ in samples 1 and 2 as $n_{g,1}$ and $n_{g,2}$, respectively, the logarithmic fold change of sample 2 compared to sample 1 is defined as $x_g(1\to2):=\log_2 n_{g,2}/n_{g,1}$ (we use ``gene count'' as a generic term for data from a proteomic/phosphoproteomic/transcriptomic measurement). To avoid an undefined expression when either of the counts is zero, a small prior count $n_0$ is usually added to both of them, so that the logarithmic fold change becomes
\begin{equation}
\label{lfc}
x_g(1\to2):=\log_2\frac{n_{g,2} + n_0}{n_{g,1} + n_0}.
\end{equation}
We use $n_0=1$ through the paper. However, this approach is not effective in proteomic data where missing values appear also in comparisons where the other count is very large. The resulting logarithmic fold change, though not infinite, is then still large and has the potential to distort the statistical analysis of the data. The different patterns of missing values in various datasets are illustrated by Fig~\ref{fig:data_properies} which shows the distributions of positive values in pairwise comparisons involving zero and a positive count. The distributions are shown for two different datasets: A transcriptomic dataset analyzed in~\cite{nisa2018comprehensive} and a phosphoproteomic dataset analyzed in [Koch et al, manuscript in preparation] (see \nameref{sec:Lluis_data} and \nameref{sec:Jonas_data} for the datasets' descriptions). To allow for easy comparison, the counts are scaled by the dataset's median in both cases. While in the transcriptomic data, zeros occur in comparisons with small counts (and hence bulk of the shown distribution is close to zero), the phosphoproteomic data are very different and contain a large number of proteins whose count changes from a large value (larger than, for example, the median count) to zero or vice versa. Missing values involved in such comparisons are referred to \emph{irregular zeros} from now on. Another way of looking at irregular zeros is provided by the following probabilistic statement: For counts greater than the dataset's median in the phosphoproteomic dataset, the probability that the other value in the comparison is zero is $0.11$. The same probability is $4.6\cdot 10^{-5}$ in the transcriptomic dataset.

When irregular zeros are present in the data, $n_0$ necessary to shrink the logarithmic fold change values computed with Eq.~(\ref{lfc}) is prohibitively large (of the order of the median count): It would significantly shrink also the logarithmic fold change values in comparisons without irregular zeros, and the differential expression analysis would be thus still distorted towards comparisons involving irregular zeros. In summary, setting $n_0>0$ is useful to shrink the logarithmic fold changes for low counts towards zero, but $n_0$ itself cannot solve the problem of irregular zeros.

Methods aiming specifically at the analysis of proteomic data acknowledge the problem of missing values and deal with it by various ``imputation'' techniques. For example, the authors of the Perseus computational platform for proteomic data~\cite{tyanova2016perseus} suggest to replace the missing values with values drawn from the dataset's empirical distribution which in addition is to be scaled and shifted. While scaling is said to prevent the imputed values from having high weight in the subsequent statistical evaluation, down-shifting is motivated by the fact that low-expression proteins are more likely to remain undetected and thus lead to zero measured counts.

\begin{figure}
\centering
\includegraphics[scale=0.85]{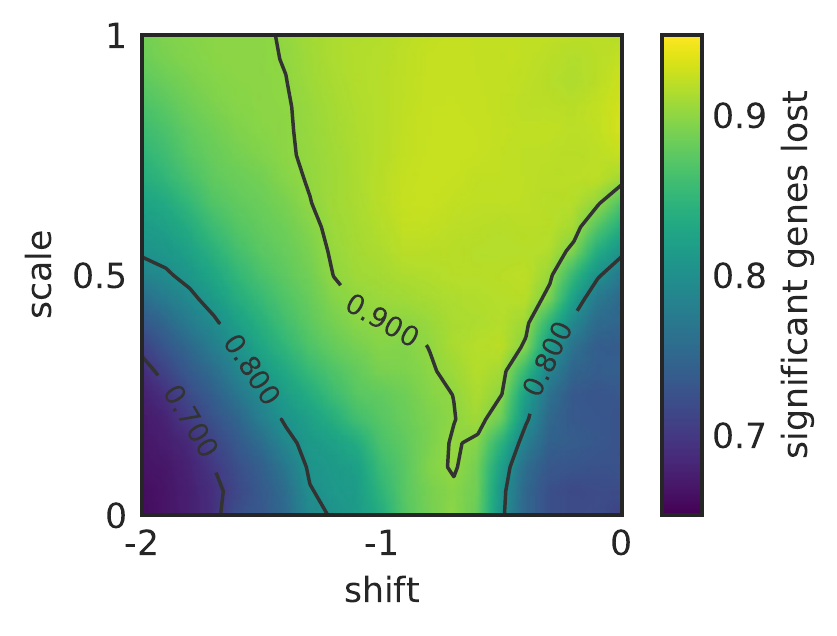}
\caption{\textbf{The impact of imputation on differential expression analysis.} The fraction of the significant genes identified by edgeR in the original transcriptomic data that are lost when the data are analyzed by edgeR after imputing missing values with various shift and scale parameters. Significance was evaluated with edgeR using the false discovery rate (FDR) threshold of 0.20. Results are averaged over 30 imputation realizations.}
\label{fig:imputation_impact}
\end{figure}

Since there are typically many missing values that need to be imputed, it is not surprising that their imputation has the potential to dramatically change results of the subsequent differential expression analysis. To demonstrate that, we apply imputation with various values of the scale and shift parameters (see \nameref{sec:imputation} for details) to the transcriptomic data which, as shown in Fig~\ref{fig:data_properies}, are free of irregular zeros and hence the results obtained with usual tools can be considered as reliable. In particular, we compare the significant genes identified by the edgeR package~\cite{robinson2010edger} in the original transcriptomic dataset with the significant genes identified by edgeR after all missing values are imputed. Fig~\ref{fig:imputation_impact} shows that the impact of imputation is substantial: Around 80\% of the originally identified significantly changing genes are lost by the recommended shift $-1.8$ and scale $0.5$. While comparatively better results are obtained with scale $0$ and shift either $-2$ or $0$ (the bottom left and right corners, respectively), around 70\% of the original significant genes are still lost. Of the two ``favorable'' settings, we thus use imputation with zero shift and scale one from now on (\ie, the missing values are replaced with the dataset's mean). The other setting, shift $-2$ and $0$, while comparatively well-performing in this evaluation, is sensitive to irregular zeros (results not shown).

\subsection*{Robustness of differential expression analysis methods to irregular zeros}
Having seen that the often-used imputation of missing values does not provide a definitive answer to the problem of irregular zeros, our main contribution is to propose a ranking-based approach to find significantly changing genes. By contrast to R's RankProd package~\cite{breitling2004rank,RankProdPackage}, which is also based on ranking the genes by fold-change values, we rank comparisons that involve a missing value separately from those that do not involve a missing value (see \nameref{sec:new_method} for a detailed description). The need for a double ranking is motivated by the fact that, as shown in \nameref{sec:missing_values}, it is impossible to rely on logarithmic fold changes when irregular zeros are present. At the same time, the missing values are not necessarily a manifestation of an erroneous or noisy measurement and can contain useful information: If a gene goes from a positive value to zero consistently in multiple comparisons, this may be an indication that the gene is downregulated in the comparison. We propose a method that takes both cases into account: The logarithmic fold changes and their magnitude relatively to other genes' fold changes are computed for comparisons without missing values, all comparisons where a zero count changes in a positive count are assigned the same relatively high virtual rank, and all comparisons where a positive count changes in a zero count are assigned the same relatively low rank. Since this method is rank-based and aims in particular at proteomic and phosphoproteomic data where the missing values problem is particularly common, we call the new method ProtRank.

To demonstrate the new method's robustness with respect to irregular zeros, we compare it with the behavior of edgeR which is probably the most popular tool for differential expression analysis~\cite{robinson2010edger} that we use both with and without imputation of missing values. We apply the methods on the transcriptomic dataset used in the previous section which, as we have seen, is essentially free of irregular zeros. We introduce the irregular zeros in the dataset by choosing at random a given fraction of positive values in the original dataset and changing them in zeros; in this way, we obtain perturbed datasets. We aim to study how does the noise in the particular form of zeros introduced in the data at random influence each respective method: EdgeR without imputation, edgeR with imputation, and the newly introduced ProtRank.

Since the original transcriptomic dataset is essentially free of irregular zeros, the significant genes identified by edgeR in the original dataset provide a natural benchmark against which results obtained with other methods can be compared; this set of original significant differentially expressed (DE) genes is denoted $\mathcal{O}$. We denote the set of significant DE genes identified by method $m$ in perturbed data as $\mathcal{P}_m$ and compare it with the original set $\mathcal{O}$. For this comparison, we use \emph{precision} and \emph{recall} which are metrics commonly used in data mining literature~\cite{manning2010introduction,lu2012recommender}. Precision is defined as the fraction of the perturbed significant genes that are also original significant genes, $\abs{\mathcal{P}_m\cap\mathcal{O}}/\abs{\mathcal{P}_m}$. Recall is defined as the fraction of originally identified significant genes that are also among the perturbed DE genes,  $\abs{\mathcal{P}_m\cap\mathcal{O}}/\abs{\mathcal{O}}$. Both metrics range from 0 (worst result) to 1 (best result). We use the described approach to evaluate the original edgeR package, edgeR with imputation, and the newly developed ProtRank method. In each case, we use the false discovery rate (FDR) threshold of $0.20$ to decide whether a gene is significant or not. To make the impact of irregular zeros explicit, we also assess the fraction of zero counts corresponding to the identified significant DE genes.

\begin{figure}
\begin{adjustwidth}{-2in}{0in}
\centering
\includegraphics[scale=0.7]{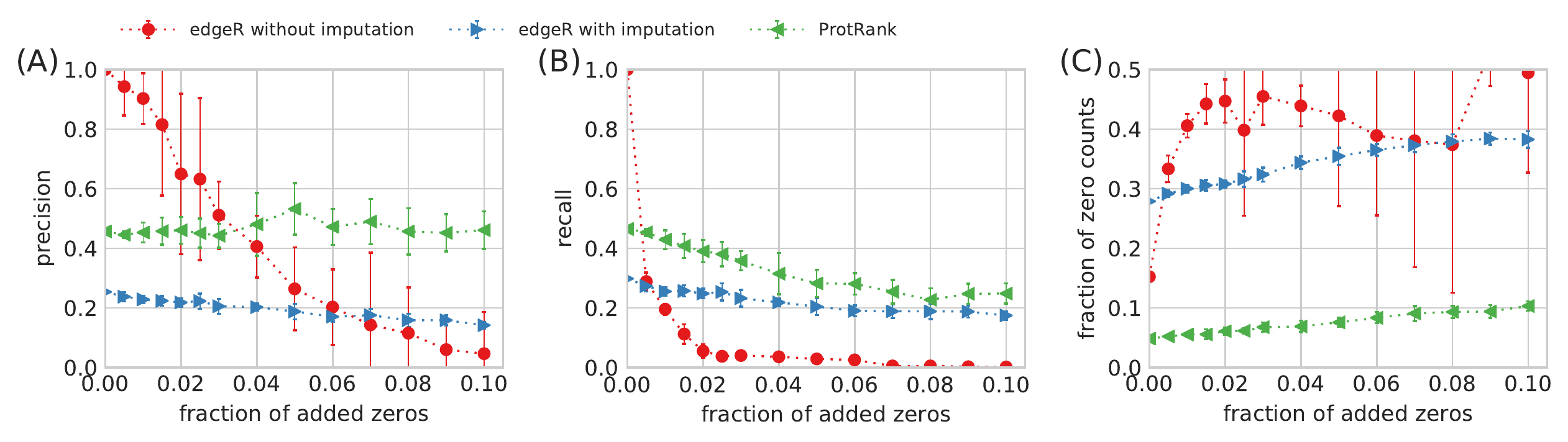}
\caption{\textbf{The impact of zeros added at random in the data.}
Precision (A) and recall (B) of various differential expression analysis methods computed with respect to the significant DE genes identified by edgeR in the original data. The fraction of zero counts among the identified DE genes (C) shows how much is each method influenced by the added zeros. The error bars show the standard deviation values computed from the analysis of 10 independent perturbed sets for each fraction of added zeros.}
\label{fig:sensitivity-zeros}
\end{adjustwidth}
\end{figure}

Fig~\ref{fig:sensitivity-zeros} summarizes the results of the robustness analysis. EdgeR without imputation naturally obtains the perfect result on unperturbed data (zero fraction of added zeros) as the benchmark DE genes are also obtained with edgeR on unperturbed data. However, its sensitivity to irregular zeros is high: When as few as 1\% of positive counts are turned into zeros, recall drops under 20\% (\ie, more than 80\% of the originally identified significant genes are lost). The method's precision decays slower but still much faster than is the case for the other methods. By contrast, edgeR with imputation is quite resistant to additional irregular zeros as its precision and recall decrease by roughly 40\% when as much as 10\% of zeros are added in data at random. Albeit stable, the results are quite bad with precision and recall decreasing from 0.31 to less than 0.20 (this is in agreement with Fig~\ref{fig:imputation_impact}). ProtRank is even more robust to irregular zeros: Its precision is stable and its recall decreases by roughly 40\% similarly to edgeR with imputation. The most important observation is that precision and recall achieved by ProtRank are significantly better than that of edgeR with imputation in the whole range of perturbation fractions. ProtRank outperforms edgeR without imputation in terms of recall (which is the more important of the two metrics as it quantifies how many of the originally found DE genes do we still find in the perturbed data) for all perturbation fractions except for the two smallest ones.

The last panel shows that the significant genes chosen by ProtRank have the smallest fraction of zero counts of the three methods. EdgeR without imputation is expectedly sensitive to the introduced zeros and the chosen significant genes have more than 40\% of zero counts when as few as 1\% of positive counts are changed in zeros. This shows that the irregular zeros, that we introduce at random and without any relation to differential expression of genes, chiefly determine which genes are chosen by edgeR as significantly differentially expressed. While less sensitive to the fraction of added zeros itself, edgeR with imputation also chooses significant genes with many zero counts (that are in turn changed in positive values by imputation). This high starting value shows that the imputation process itself, albeit assumed to solve the problem of missing values, biases the selection of significantly DE genes towards the genes that have many missing values. By contrast, ProtRank chooses significant genes with few zero counts and the fraction of zero counts increases slowly with the fraction of added zeros. 

\begin{figure}
\begin{adjustwidth}{-2in}{0in}
\centering
\includegraphics[scale=0.7]{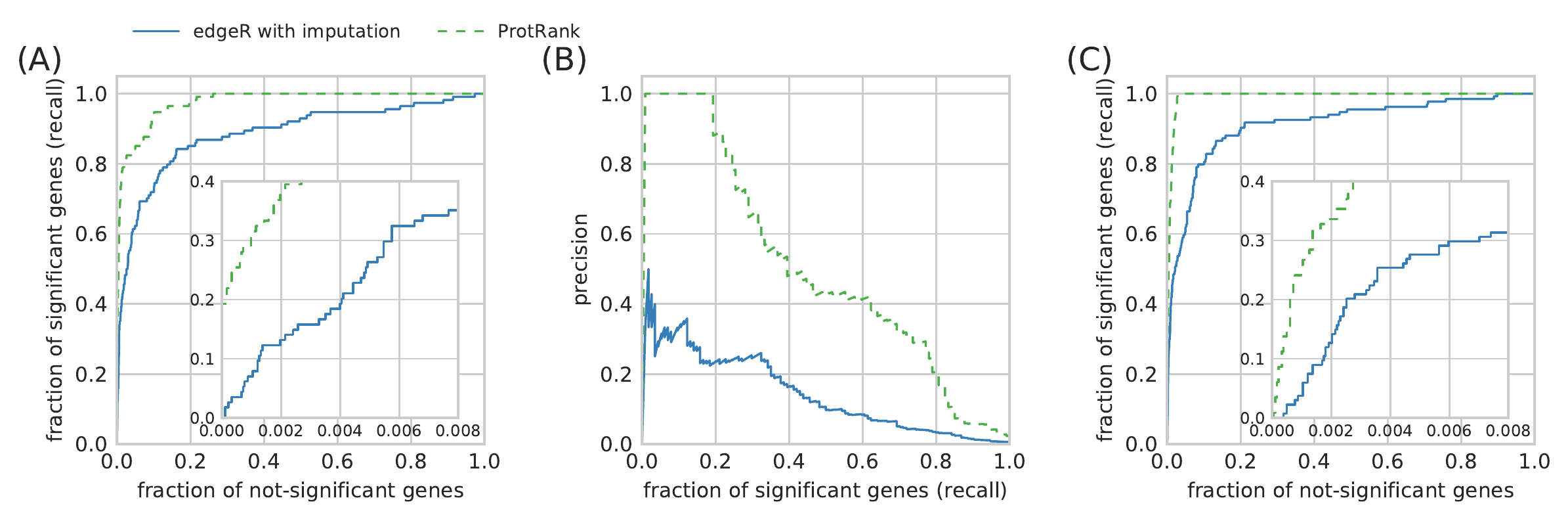}
\caption{\textbf{A comparison of the significant genes and rankings obtained with respective methods.} (A) The receiver operating characteristic (ROC) curves based on comparing with the DE genes identified by edgeR; the bottom-left corner is magnified in the inset. (B) The precision-recall (PR) curves based on comparing with the DE genes identified by edgeR. (C) The receiver operating characteristic (ROC) curves based on comparing with the DE genes identified by ProtRank and edgeR with imputation, respectively, in the ranking of genes produced by edgeR on the original data.}
\label{fig:ROC}
\end{adjustwidth}
\end{figure}

To better understand the difference between the results produced by the three considered methods, we evaluate the positions of the DE genes identified by edgeR in the other two rankings: The ranking produced by edgeR with imputation and the ranking produced by ProtRank (in the rankings, the genes are ranked by the significance of their differential expression from the most to the least significant). In Fig~\ref{fig:ROC}, we visualize the comparison using the well-known receiver operating characteristic (ROC) curve~\cite{fawcett2006introduction,davis2006relationship} and the precision-recall (PR) curve that has been advocated for use in biological data in~\cite{chicco2017ten}. The ROC curve in Fig~\ref{fig:ROC}A, in particular, the inset focusing at the top of the rankings, show that the ProtRank's ranking has the edgeR's DE genes at higher positions than edgeR with imputation does.

The reason why \cite{chicco2017ten} suggest to use the PR curve instead of the ROC curve is that the number of \emph{positive instances} (in our case represented by the correctly identified differentially expressed genes) is much smaller than the number of \emph{negative instances} (in our case represented by the genes that are correctly identified as not differentially expressed). The ROC curve involves \emph{true negatives} in its computation which, due to their abundance, give rise ROC curves that have a large area under them; this area is a common way to quantify a ROC curve. This is well visible in panel Fig~\ref{fig:ROC}A where the areas under the two ROC curves are 0.98 and 0.87, respectively. The PR curves in Fig~\ref{fig:ROC}B overcome this limitation and make a clear distinction between the two evaluated methods: The area under the ProtRank's PR curve, 0.45, is four times as large as the area 0.11 produced by edgeR with imputation.

We finally do a reverse check and examine the positions of the significant DE genes identified by edgeR with imputation and ProtRank, respectively, in the ranking of genes by the significance of their differential expression produced by edgeR. The result is shown in Fig~\ref{fig:ROC} where it is immediately visible that the significant genes chosen by ProtRank are all highly ranked in the original gene ranking produced by edgeR without imputation. In fact, all ProtRank's significant genes (we use the FDR threshold of $0.20$ again) are in the top 2.2\% of the ranking of genes by edgeR in the unperturbed data. By contrast, a substantial fraction of genes chosen by edgeR with imputation are scattered through the lower parts of the original gene ranking. This shows that in the absence of irregular zeros, results obtained with ProtRank are similar to those obtained with edgeR without imputation of missing values.

\subsection*{Results on the phosphoproteomic data}
\label{sec:phospho_results}
We now return to the phoshoproteomic data that initially motivated the development of the statistical framework that we introduce in this paper (see \nameref{sec:Jonas_data} for a detailed dataset description). Since the complete analysis of the data, supplemented by extensive biological experiments, will be part of a forthcoming manuscript [Koch et al, manuscript in preparation], we provide here only a general evaluation of the identified significantly differentially expressed phosphopeptides. Without the loss of generality, we use the data corresponding to four samples measured one hour after irradiation (SAMT\_IR1h\_1, SAMT\_IR1h\_2, SAYH\_IR1h\_1, SAYH\_IR1h\_2) and four corresponding controls (SAMT\_C\_1, SAMT\_C\_2, SAYH\_C\_1, SAYH\_C\_2). As explained in \nameref{sec:Jonas_data}, SA, MT, and YH are three respective mutations that the sample cells can have; labels 1 and 2 mark two biological duplicates that were available for each mutation combination. In the measurement data, there are 6201 peptides that have at least one positive count in the eight aforementioned samples. The dataset contains $31\%$ of zero counts, many of which are irregular zeros (\ie, they correspond to a pair of samples where the given phosphopeptide's count in the other sample is larger than the median count).

The numbers of DE phosphopeptides identified by respective approaches using the FDR threshold 0.20 are: 1278 for edgeR without imputation, 93 for edgeR with imputation, and 45 for ProtRank. The result obtained with edgeR without imputation is clearly excessive with more than 20\% of all peptides being identified as significantly differentially expressed. This is due to the irregular zeros that distort the results; this is shown by 63\% of the DE phosphopeptides' counts being zeros, which is more than double of the overall fraction of zero counts in the data. EdgeR with imputation does not have a similar problem and yields a similar number of differentially expressed phosphopeptides as ProtRank.

To gain further insights, we evaluate median counts (computed from the positive counts only) of the identified DE phosphopeptides, in particular in comparison with median counts of all phosphopeptides. Denoting the fraction of the identified DE phosphopeptides in the count bin $b$ as $f^{DE}_b$ and the fraction of all phosphopeptides in the count bin $b$ as $f^0_b$, the ratio $f^{DE}_b/f^0_b$ quantifies the relative representation of DE phosphopeptides from the given count bin $b$. When the relative representation is more than one, the given count bin $b$ is \emph{over-represented} among the identified DE phosphopeptides. When the relative representation is less than one, the given count bin $b$ is \emph{under-represented} among the identified DE phosphopeptides.

\begin{figure}
\centering
\includegraphics[scale=0.7]{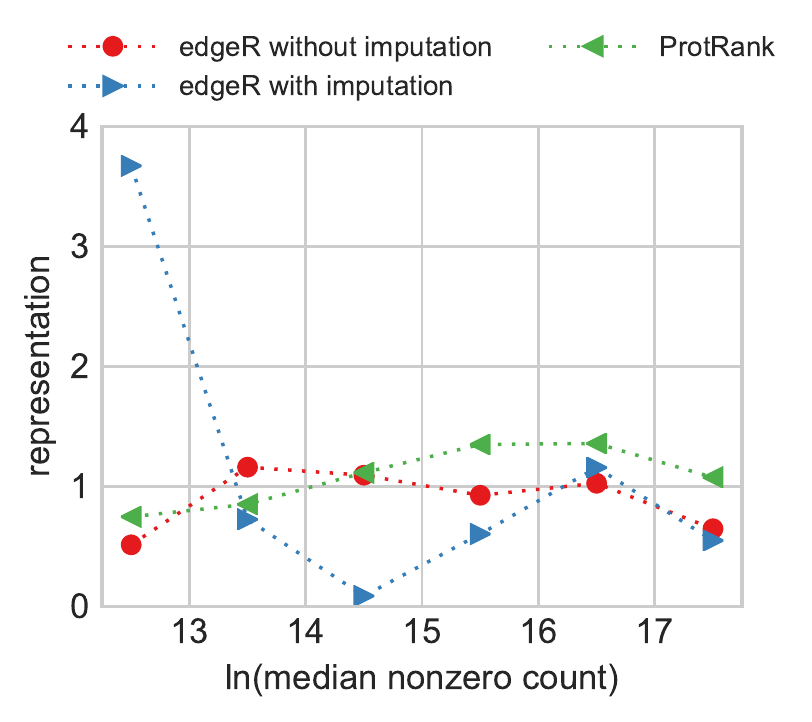}
\caption{\textbf{The relative representation of DE phosphopeptides identified by various methods as a function of their log-transformed median count.} The representation value of one indicates that phosphopeptides with the given median count are neither over-represented nor under-represented among the identified DE phosphopeptides.}
\label{fig:phosphoprot}
\end{figure}

The result is shown in Fig~\ref{fig:phosphoprot} which shows that the three evaluated approaches greatly differ in how their representation changes with the median phosphopeptide count. While edgeR without imputation and ProtRank show little bias over the whole range of median counts, edgeR with imputation shows a strong bias against phosphopeptides whose counts are close to the overall average count. EdgeR's behavior is a direct consequence of the imputation process that replaces missing values with mean count (in our case) and thus makes it possible that the phosphopeptides with low or high median count can have high apparent changes between their low/high actual counts and the average counts introduced by imputation. This is well visible in Fig~\ref{fig:phosphoprot} where bins close to the average count are strongly under-represented, and bins containing phosphopeptides with low/high counts are over-represented.

To summarize the results obtained on the phosphoproteomic data: ProtRank yields a plausible number of DE phosphopeptides which furthermore show no systemic biases. By contrast, edgeR without imputation produces an excess number of DE phosphopeptides and edgeR with imputation is strongly skewed toward phosphopeptides that have either low or high counts.

The complex experimental setup of this dataset allows us to illustrate another ProtRank's asset: The possibility to simultaneously address all possible differential expression patterns (see \nameref{sec:new_method} for details). In the case of the given dataset, for example, it is possible that the two types of samples (SAMT and SAYH) react to irradiation in a different way: Some phosphopeptides can increase upon irradiation in SAMT samples and decrease upon irradiation in SAYH samples, for example. Besides the baseline comparison of all irradiated samples with their counterpart controls, ProtRank makes it possible to consider two separate groups---SAMT samples and SAYH samples, in this case. The rank score of each gene is then computed in such a way that the genes that consistently increase upon irradiation in both SAMT and SAYH samples, the genes that consistently decrease upon irradiation in both SAMT and SAYH samples, as well as the genes that increase upon irradiation in SAMT and decrease in SAYH samples (and vice versa), are assigned a high rank score. From the 45 genes identified by ProtRank as differentially expressed, 18 increase upon irradiation in both SAMT and SAYH, 18 decrease in both SAMT and SAYH, 6 increase in SAMT and decrease in SAYH, and 3 decrease in SAMT and increase in SAYH. The increase in SAMT and decrease in SAYH combination produces a particularly robust signal with two of the identified genes having $\text{FDR} < 0.01$.

Similar multi-directional analyses are also possible using other tools---such as edgeR that we use here for comparison---but they need to be manually done separately for each combination of directions, and the obtained results have to be compiled whilst explicitly taking into account that they come from multiple comparisons. ProtRank allows the same analysis to be carried out in two lines (first defining the groups of comparisons to be made, second calling ProtRank's main gene-ranking function).

\subsection*{Comparison with Perseus and Proteus}
\label{sec:PerProt}
We have shown so far that ProtRank overcomes the problem of missing values without the need to impute them and, at the same time, produces results that are in the absence of irregular zeros to a large extent comparable with results produced by the state-of-art differential analysis tool edgeR~\cite{robinson2010edger}. We now proceed by briefly comparing the ProtRank results with the results obtained by two other existing tools: (1) Perseus, a well-established computational platform for proteomic data~\cite{tyanova2016perseus} which uses imputation to deal with missing values and (2) Proteus, a recent R package for the analysis of quantitative proteomics data~\cite{gierlinski416511}. As we further argue in Discussion, a comprehensive comparison of the available tools should also include the use of synthetic datasets benchmarked against a number of different real datasets. We see this as an important task for future research.

With Perseus, we used the transcriptomics dataset which, as we have seen, is essentially free of irregular zeros and EdgeR is therefore expected to produce a meaningful differential expression analysis result. We first log-transformed the counts and then imputed the missing values~\cite{tyanova2016perseus}. While the software offers two different imputation approaches, the default imputation based on the mean and the standard deviation for each sample and the ``global'' imputation based on the mean and the standard deviation for all samples, the results are qualitatively similar for both of them. We used the function ``multiple-sample test'' to identify the genes that are differentially expressed between the primary and recurrent tumor samples. Despite trying various analysis settings, Perseus identifies a small number of genes as differentially expressed. We thus focus our comparison on the top 100 most differentially expressed genes, even when they are not marked as significant. Of them, less than 10\% are among the 114 significantly DE genes identified by edgeR. The area under the precision-recall curve is 0.04 as opposed to 0.45 achieved by ProtRank. In other words, the ranking of genes produced by Perseus substantially differs from the ranking of genes produced by edgeR. This is in line with our previous observation that imputation has the potential to dramatically alter the differential expression analysis results. By changing a chosen fraction of counts to zeros, we can further probe the Perseus's robustness with respect to artificially introduced zeros. When 0.02 of all counts are chosen at random and changed to zeros, 40\% of the original top 100 genes remain in the top 100 on average. When the fraction of zeros introduced at random increases to 0.10, 25\% of the original top 100 genes remain in the top 100 on average. These results are similar to those achieved by ProtRank.

Proteus is designed to use an evidence file from MaxQuant as input and currently lacks the possibility to use a simple table of peptide intensities instead; to apply Proteus on the datasets studied here so far is therefore not possible. Instead, we use the peptide intensities produced by the example described in~\cite{gierlinski416511} in Section 2.1. This dataset contains data on 34,733 peptides measured in two biological conditions, named A and B, and seven replicates each (14 samples in total). The dataset has the properties expected for a mass spectrometry proteomics measurement: 26\% of all counts are zeros and 6.5\% of all A vs. B comparisons involve irregular zeros. EdgeR is not expected to produce high quality results for such input data. Proteus and ProtRank yield similar numbers of significant DE peptides, 317 and 344, respectively, for the significance threshold of 0.05, for example. The overlap between these two sets of DE peptides is 120 with further 100 peptides in each set identified as DE by the other tool at the significance level of 0.20. Similar findings follow on the aggregated protein intensities data obtained with the Proteus's function \texttt{makeProteinTable} (the resulting dataset contains 3525 proteins). Proteus and ProtRank then yield 42 and 76 significant DE proteins, respectively, at the significance threshold of 0.05, and the overlap of these two sets is 29. The conclusion is that the DE analysis results obtained with Proteus and ProtRank are similar.

\section*{Discussion}
We have shown that the presence of irregular zeros---missing values that in the differential expression (DE) analysis occur in comparisons with substantial measured values---in proteomics data importantly influences the resulting lists of DE genes produced by common statistical tools such as edgeR. We stress that for other kinds of input data, such as transcriptomics data, edgeR is a good statistical tool that produces high-quality results.

There are two main directions that we see for the future development of ProtRank. Firstly, the computationally intensive bootstrap step (which is used to estimate the false detection rate, FDR) could be replaced by an approximate analytical procedure. The first motivation for such an approach is provided by Fig~\ref{fig:bootstrap} where the aggregate bootstrap scores decay exponentially at the top positions of the gene ranking. Analytical estimates of the bootstrap score distribution could then be used as a replacement for the actual bootstrap procedure.

Secondly, a different recent method, Proteus, addressing the problem of missing values in proteomics~\cite{gierlinski416511} came to our attention in the final stages of this manuscript's preparation. We have shown in \nameref{sec:PerProt} that albeit similar, the results produced by ProtRank and Proteus differ in the evaluation of numerous peptides and proteins. A detailed comparison of these methods on various proteomic and phosphoproteomic datasets as done, for example, in~\cite{soneson2013comparison} for methods designed for transcriptomic data, is the natural first step. The natural limitation of such a comparison is that the ground truth (the ``correct'' list of differentially expressed peptides or phosphopeptides) is not known. This can be alleviated by evaluating the methods also on synthetic datasets. Upon careful calibration, synthetic datasets can share many of real datasets' features which makes the subsequent evaluation of methods more credible~\cite{medo2016model}. Combined benchmarking of methods on real and synthetic datasets could help establish a comprehensive robust statistical framework for the analysis of proteomic data.

\section*{Conclusions}
We propose here a novel method for differential expression analysis of proteomic and phosphoproteomic data. The main advantage of this new method is that it is robust to the missing values that are common for proteomic and phosphoproteomic measurements. As a result, it does not require the imputation step which is commonly used to eliminate the missing values~\cite{tyanova2016perseus}, yet we show here that it at the same time importantly affects the obtained results. In data where missing values are absent, the new method---which we refer to as ProtRank because it is based on rankings---produces similar results as edgeR which is a widely-used method for differential expression analysis. When missing values are artificially introduced in the data, ProtRank's results are more stable than the results produced by edgeR which is a demonstration of ProtRank's robustness to missing values.

ProtRank requires no parameters to be fine-tuned for the analysis. It also does not employ any normalization of counts in individual samples as this would not change the gene ranking that is taken into account by the method (only the numeric fold change values would change upon normalization). Importantly, ProtRank makes it possible to automatically address more complex differential expression patterns such as the case discussed in \nameref{sec:phospho_results} where irradiation was applied on samples with various mutations and it was, in principle, possible that samples with one mutation react to irradiation differently than samples with other mutations. An implementation of ProtRank is available at \url{https://github.com/8medom/ProtRank} as an easy to use Python package.

\section*{Declarations}

\subsection*{Ethics approval and consent to participate}
Not applicable.

\subsection*{Consent for publication}
Not applicable.

\subsection*{Availability of data and material}
The datasets analyzed in this manuscript are available from the corresponding author on reasonable request.

\subsection*{Competing interests}
The authors declare that they have no competing interests.

\subsection*{Funding}
This work has been supported by Stiftung für klinisch-experimentelle Tumorforschung (grant to MiMe).

\subsection*{Authors' contributions}
DMA and MiMe conceived the project. MaMe developed the methodology and analyzed the results. MaMe wrote the manuscript. DA provided administrative and technical support. All authors read and approved the final manuscript.

\subsection*{Acknowledgements}
We thank Jonas Koch from the Radiation Oncology group at Inselspital/University of Bern and Eleonora Orlando from the Quantitative Mass Spectrometry group at ETH Z\"urich for providing us the proteomic data. We thank Lluís Nisa \emph{et al} for providing us the transcriptomic data.

\section*{Methods}
\label{sec:methods}

\paragraph*{Description of the transcriptomic data}
\label{sec:Lluis_data}
The transcriptomic data from head and neck squamous cell carcinoma patients have been originally analyzed in~\cite{nisa2018comprehensive} (the authors have used edgeR for the differential expression analysis). Out of the 15 cell lines used in that study, we keep eight of them for pairwise differential expression analysis: Cell lines UM-SCC-11A, -14A, -74A, and -81A from primary tumors, and cell lines UM-SCC-11B, -14B, -74B, and -81B from recurrent tumors. This corresponds to four pairwise comparisons (recurrent vs. primary) in total. Each sample has been measured once with RNA sequencing which produced integer counts of 18,369 distinct transcripts (see \cite{nisa2018comprehensive} for details of the experimental setup). In the data, $11.9\%$ of all counts are zero and the median of positive counts is $932$. The fraction of pairwise comparisons where one count is greater than this median and the other count is zero, is less than 0.01\% (3 comparisons out of 35,359); irregular zeros are thus essentially absent in this dataset.

\paragraph*{Description of the phosphoproteomic data}
\label{sec:Jonas_data}
NIH3T3 mouse embryonic fibroblasts were transfected with either the wild-type or one of five mutated forms (M1268T, Y1248H, S1014A, M1268T/S1014A, Y1248H/S1914A) of the MET receptor tyrosine kinase were irradiated with a single dose of 10 Gy (Gammacell GC40, MDS Nordion, Ontario, Canada). Samples from the six cell lines have been subjected to phosphoproteomic analysis via non-targeted mass spectrometry before irradiation, 1 hour after irradiation, and 7 hours after irradiation. Since two biological duplicates of each of the cell lines have been analyzed, the phosphoproteomic results are available for $6\times3\times2=36$ samples in total. Integer peptide counts of 7,572 unique peptides are available for each sample. In the data, $43.6\%$ of all counts are zero and the median of positive counts is $1,294,600$. The fraction of pairwise comparisons where one count is greater than this median, yet the other count is zero is $10.8\%$; irregular zeros are frequent in this dataset. An in-depth analysis of this dataset will be presented in [Koch et al, manuscript in preparation].

\paragraph*{Phosphoproteomic data experimental setup}
Cell cultures were washed, scraped in phosphate-buffered saline and spun down for 5 minutes at 1000\,rpm. Resulting pellets were resuspended in 8\,M urea solution containing 0.1\,M ammonium bicarbonate and disrupted by sonication. Supernatants were centrifuged at 12000\,rpm for 10 minutes and protein concentration was determined by BCA Protein Assay (Pierce). Disulfide bonds were reduced with tris(2-carboxyethyl)phosphine at a final concentration of 5\,mM at 37$^\circ$C for 30 minutes and alkylation of free thiols was performed with 10\,mM iodoacetamide at room temperature for 30 minutes in the dark. The solution was subsequently diluted with 0.1\,M ammonium bicarbonate to a final concentration of 1.5\,M urea and digestion was performed overnight at 37$^\circ$C by sequencing-grade modified trypsin (Promega) at a protein-to-enzyme ratio of $50:1$. Acidification was performed by adding formic acid to a final $\text{pH}<3$ in order to stop protein digestion. Peptides were desalted on a C18 Sep-Pak cartridge (Waters) and one-tenth of the resulting eluate was processed individually for total proteome analysis.  Phosphopeptides were enriched from 1\,mg of initial peptide mass with TiO$_2$ as previously described~\cite{bodenmiller2007reproducible}. For mass spectrometry analysis, samples were resuspended in 20$\,\mu$l of 2\% acetonitrile, 0.1\% formic acid, and 1$\,\mu$l of each sample was used for injections. LC-MS/MS analysis was performed with an Easy nLC 1000 system (Thermo) connected to an Orbitrap Elite mass spectrometer (Thermo) equipped with a NanoFlex electrospray source. Peptides were separated on an Acclaim PepMap RSLC C18 column (150\,mm $\times$ 75$\,\mu$m, 2\,um particle size, Thermo) using a gradient of 5--30\% buffer B (98\% acetonitrile, 2\% water, 0.15\% formic acid) over 180 minutes at a flow rate of 300\,nl/min. The Orbitrap Elite was operated in data-dependent acquisition mode, each cycle consisting of one MS scan followed by 15 MS/MS scans of the most abundant precursor ions. Collision-induced dissociation was performed with the following settings: Isolation width, 2\,m/z; normalized collision energy, 35; activation time, 10\,ms. Acquired MS data files were subsequently processed for identification and quantification using Maxquant version 1.5.2.8~\cite{cox2008maxquant}. Settings were kept as default with the following specifications: 'First search peptide tolerance' was set to 50\,ppm and 'Main search peptide tolerance' to 10\,ppm. The considered modifications were oxidation (Met) and phosphorylation (Ser/Thr/Tyr). 'Label free quantification' and 'Match between runs' were enabled, with a match time window of two minutes. The search was performed against the mouse UniProt FASTA dataset UP000000589.

\begin{figure}
\centering
\includegraphics[scale=0.7]{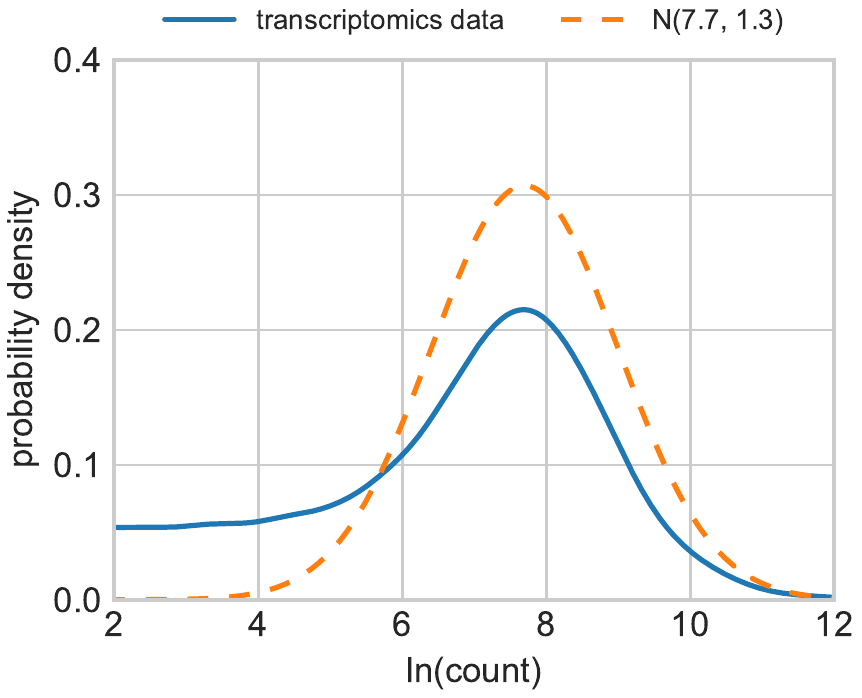}
\caption{Distribution of the logarithm of the positive counts in the phoshoproteomic data (solid line) and the normal distribution with $\mu_0=7.7$ and $\sigma_0=1.3$ that fits the bulk of the distribution.}
\label{fig:imputation_parameters}
\end{figure}

\paragraph*{Imputation of missing values}
\label{sec:imputation}
As can be seen in Fig~\ref{fig:imputation_parameters}, the bulk of the distribution of the logarithm of positive gene counts can be well fitted with the normal distribution with mean $\mu_0=7.7$ and standard deviation $\sigma_0=1.3$. In line with~\cite{tyanova2016perseus}, we thus replace the missing values with $\exp(V)$ where $V$ is drawn from the normal distribution with mean $7.7 + \delta \sigma_0$ and standard deviation $\lambda\sigma_0$. The exponential transformation is needed here to go from the logarithmic counts used for display in Fig~\ref{fig:imputation_parameters} back to the natural range and scale of gene counts. Parameters $\delta$ and $\lambda$ are referred to as shift and scale, respectively. While~\cite{tyanova2016perseus} recommends the choice $\delta=-1.8$ and $\lambda=0.5$, the investigation of the parameter space in Fig~\ref{fig:imputation_impact} suggests $\delta=0$, $\lambda=0$ to be a better choice when imputed data are used as input for the edgeR package and analyzed using its functions \verb|calcNormFactors|, \verb|EstimateDisp|, \verb|glmFit|, and \verb|glmLRT|.

\paragraph*{ProtRank description}
\label{sec:new_method}
Counts $n_{g, i}$ of all genes ($g=1,\dots,G$; we use ``gene'' here as a generic term here) in all samples ($i=1,\dots,S$) serve as input data for the analysis. The samples are organized in $C=S/2$ pairwise comparisons ($c=1,\dots,C$) that represent a change of conditions (before and after treatment, typically) for a given system (a cell line or a patient). The comparisons are further organized in groups of comparisons that are expected to go in the same direction. In the analysis described in \nameref{sec:phospho_results}, one group of comparisons includes SAMT samples and the other group of comparisons includes SAYH samples. If the change is expected to be in the same direction in all samples (as would be the case when analyzing data from biological or technical duplicates of a system), there is only one group including all analyzed sample comparisons. The opposite extreme where each comparison forms a separate group is also possible, but groups involving more comparisons (including duplicates of the same conditions or comparisons of several samples that are expected to react uniformly) are likely to produce better results. Denoting the set of all comparisons as $\mathcal{C}$, we can write $\mathcal{C} = \{G_1, \dots, G_N\}$ where $N$ is the number of comparison groups. Then for group $G_n$ is composed of individual comparisons, $G_n=\{c_{n,1}, \dots, c_{n,M_n}\}$ where $M_n$ is the number of comparisons in group $n$. Finally, comparison $c_{n,j}$ is composed of two samples, $c_{n,j}=\{s_{n,j}^B, s_{n,j}^A\}$ which correspond to the sample before and after treatment, respectively.

We first consider a single comparison of samples $s_{n,j}^B$ and $s_{n,j}^A$. For all genes $g$ that have been registered in both these samples, we denote their number as $\Omega(s_{n,j}^B, s_{n,j}^A)$, we compute their logarithmic fold change values
\begin{equation}
x_g(s_{n,j}^B\to s_{n,j}^A)=\log_2\frac{n_{g,s_{n,j}^A} + n_0}{n_{g,s_{n,j}^B} + n_0}
\end{equation}
and consequently compute their rank $r_g(s_{n,j}^B\to s_{n,j}^A)$ by the logarithmic fold change from the highest (ranked 1) to the lowest [ranked $\Omega(s_{n,j}^B, s_{n,j}^A)$]. This rank is further rescaled to the rank score
\begin{equation}
\sigma_g(s_{n,j}^B\to s_{n,j}^A) = \frac{r_g(s_{n,j}^B\to s_{n,j}^A) - 0.5}{\Omega(s_{n,j}^B, s_{n,j}^A)}
\end{equation}
which, thanks to the shift by $0.5$, is symmetrically distributed in the range $[0, 1]$ (the lowest rank score is as far from zero as the highest is from one). The rank score $\sigma$ is the basis of ProtRank's gene ranking.

Before proceeding, we have to assign a rank score to the genes that have zero counts in either (or both) of the compared samples. Since the change from a zero count to a positive count corresponds to a large positive logarithmic fold change, we assign those pairs uniform rank score $\sigma_0$ which is the method's parameter. We set $\sigma_0=0.1$ which corresponds to assigning the change from zero to a positive count the same score as assigned to a pair of two positive counts with the 10th percentile logarithmic fold change. In general, lower values of $\sigma_0$ result in a higher fraction of zero counts among the identified differentially expressed genes. The precise choice of $\sigma_0$ is made less important by Eq.~(\ref{rank_score_start}) which log-transforms the computed scores. Analogously, the change form a positive count to zero corresponds to a large negative logarithmic fold change. We assign those pairs with score $1 - \sigma_0$ which is the same as the score assigned to a pair of two positive counts with the 90th percentile logarithmic fold change. Finally, pairs with two zero counts are ignored in the computation of the final score because they provide no useful information for the differential expression analysis.

To rank the genes based on all comparisons, we now have to aggregate individual rank scores into a final rank score. Similarly to~\cite{breitling2004rank,RankProdPackage}, this is done by multiplying the logarithm of individual rank scores from all comparison groups as
\begin{equation}
\label{rank_score_start}
\prod_{n=1}^{N}\prod_{j=1}^{M_n} \big[-\ln\sigma_g(s_{n,j}^B\to s_{n,j}^A)\big].
\end{equation}
Terms corresponding to comparisons involving two zero counts are ignored from the aggregation process. To understand why the logarithmic transformation here is preferable to directly multiplying individual rank scores, consider the case where a gene has the largest positive logarithmic fold change of all in one comparison and the largest negative logarithmic fold change in another one. Assuming that there are, for example, $N=5000$ genes in total, the respective rank scores are $0.5/N = 10^{-4}$ and $1 - 0.5 / N = 1 - 10^{-4}$, respectively. Their direct multiplication then yields approximately $10^{-4}$ which is the same values as a gene whose logarithmic fold change is 50th largest in both cases (both individual rank scores would then be approximately $10^{-2}$). This is obviously an undesired outcome as the two mentioned genes are far from being similarly differentially expressed. The problem is overcome by aggregating according to Eq.~(\ref{rank_score_start}): The first gene then scores $9.21 \times 10^{-4}\approx 10^{-3}$ which is far less than the second gene whose score is $4.6\times 4.6\approx21.2$. In other words, Eq.~(\ref{rank_score_start}) favors the genes whose expression changes similarly in all comparisons at the cost of genes whose expression changes wildly in different directions.

\begin{figure}
\centering
\includegraphics[scale=0.7]{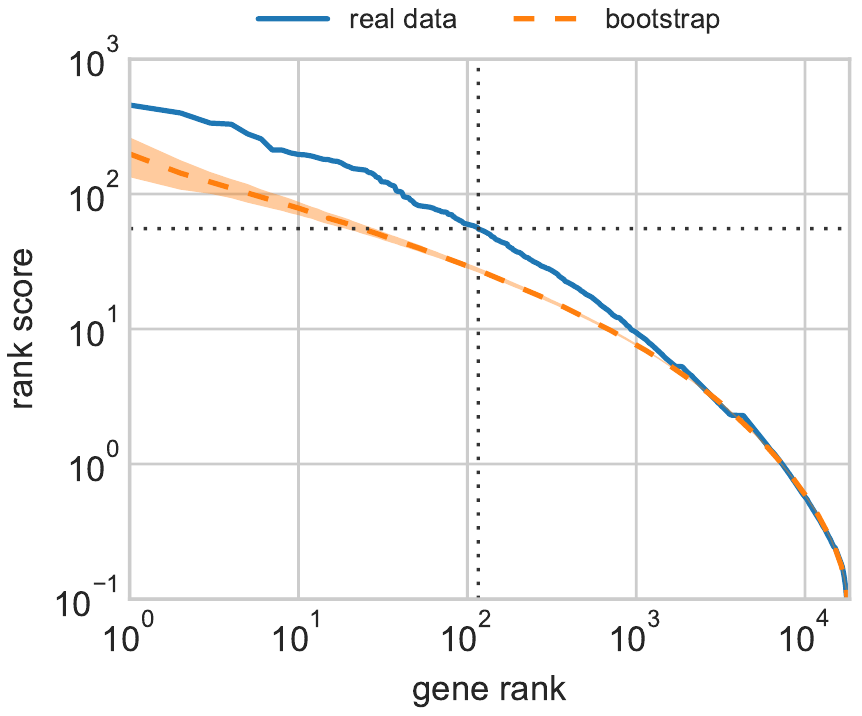}
\caption{The rank plot of the gene rank score computed by ProtRank in the transcriptomic data (real and bootstrapped data). The shaded region indicates the standard deviation in bootstrap realizations. At the FDR threshold of 0.20, ProtRank finds 116 significant DE genes. In the bootstrap data, there are 23 genes that have a better averafe score than the 116th gene in the real data ($23/116\approx 0.198$).}
\label{fig:bootstrap}
\end{figure}

Given the final score obtained with Eq.~(\ref{rank_score_start}), genes with the highest final rank score are candidates for differentially expressed genes. However, this would seek only for genes with small rank score (that is, large positive logarithmic fold change) in all comparisons. To achieve our goal of allowing to search for genes with different direction of change in respective comparison groups, we introduce the set of group directions $\mathcal{D}=\{d_1,\dots,d_N\}$, where $d_n=-1$ corresponds to searching for genes with large negative logarithmic fold change in group $n$ (\ie, high rank score $\sigma$) and $d_n=1$ corresponds to searching for genes with large positive logarithmic fold change in group $n$ (\ie, low rank rank score $\sigma$). The final rank score is then computed as the minimal rank achieved out of all possible group directions,
\begin{equation}
\label{rank_score}
\sigma_g = \min_{\mathcal{D}} \prod_{n=1}^{N}\prod_{j=1}^{M_n}
\Big[-\ln\Big(\frac12 - \frac{d_n}2 + d_n\,\sigma_g(s_{n,j}^B\to s_{n,j}^A)\Big)\Big],
\end{equation}
where the inner-most term simplifies to $\sigma_g(s_{n,j}^B\to s_{n,j}^A)$ when $d_n=1$ (seeking for genes upregulated in group $n$) and to $1-\sigma_g(s_{n,j}^B\to s_{n,j}^A)$ when $d_n=-1$ (seeking for genes downregulated in group $n$). Terms corresponding to comparisons involving two zero counts are again ignored from computing the aggregate score. Genes are then ranked by their final rank score from the highest to the lowest.

To decide which genes are significantly differentially expressed in the data, we use nonparametric bootstrap~\cite{shalizi2010bootstrap,shao2012jackknife}: We create simulated score tables by randomizing the gene rank scores for each individual comparison. The original final rank scores of the genes are then compared with their final rank scores in bootstrap realizations. This allows us to determine which rank scores in the original data are perhaps small but still likely to occur by chance, and which are so small that they correspond to differentially expressed genes (see Fig~\ref{fig:bootstrap} for an illustration). Now take gene $g$ that is ranked $r_g$ by Eq.~(\ref{rank_score}); the most differentially expressed gene has rank $1$. In each bootstrap realization, we compute the number of genes whose final rank score is better than the real final rank score of gene $g$, and compute their average number $N_g$ over all realizations. The false discovery rate for gene $g$ is then estimated as $N_g/r_g$. As one proceeds down the ranking, the quantity $N_g/r_g$ can sometimes decrease. In such a case, we assign the gene with the false discovery rate estimated for the previous better-ranked gene. This forces the estimated false discovery rate to increase monotonously.

\paragraph*{How to use the ProtRank package}
\label{sec:PR_usage}
A Python implementation of the new method can be downloaded from \url{https://github.com/8medom/ProtRank}. The github repository contains the package file ProtRank.py as well as the synthetic dataset \verb|sample_dataset.dat| and the Python script \verb|sample_dataset_analysis.py| which provides a simple example of how to use the ProtRank package.

The elementary package usage comprises two steps: loading the data for analysis using \verb|load_data| function and carrying out the differential expression analysis using \verb|rank_proteins| function. To analyze the aforementioned synthetic dataset, the minimal example is:

\lstset{basicstyle=\ttfamily\small,breaklines=true}
\begin{adjustwidth}{-1in}{0in}
\begin{lstlisting}[language=Python]
import ProtRank

what_to_compare = [[['A1', 'B1'], ['A2', 'B2'], ['A3', 'B3'], ['A4', 'B4']]]
description = 'A_vs_B'

dataset = ProtRank.load_data('sample_dataset.dat')
significant = ProtRank.rank_proteins(dataset, what_to_compare, description)
\end{lstlisting}
\end{adjustwidth}
Variable \verb|significant| stores the list of the identified differentially expressed proteins (identified by the index of the corresponding rows).

In addition, basic statistical properties of the dataset can be displayed using \verb|data_stats| function, and the logarithmic fold changes computed for selected rows (typically those corresponding to the identified differentially expressed genes; we can use the list \verb|significant| created by the code above, for example) can be visualized using \verb|plot_lfc| function.

\end{document}